# SARA: A Microservice-Based Architecture for Cross-Platform Collaborative Augmented Reality

**Diego Vaquero-Melchor** [1,2,*] **and Ana M. Bernardos** [1,2] **and Luca Bergesio** [1,2]

1. Information Processing and Telecommunications Center, Universidad Politécnica de Madrid, 28040 Madrid, Spain; abernardos@grpss.ssr.upm.es (A.M.B.); luca.bergesio@grpss.ssr.upm.es (L.B.)
2. ETSI Telecomunicación, Av. Complutense 30, 28040 Madrid, Spain
* Correspondence: diego.vaquero@grpss.ssr.upm.es

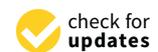

**Abstract:** Augmented Reality (AR) functionalities may be effectively leveraged in collaborative service scenarios (e.g., remote maintenance, on-site building, street gaming, etc.). Standard development cycles for collaborative AR require to code for each specific visualization platform and implement the necessary control mechanisms over the shared assets. in order to face this challenge, this paper describes SARA, an architecture to support cross-platform collaborative Augmented Reality applications based on microservices. The architecture is designed to work over the concept of collaboration models which regulate the interaction and permissions of each user over the AR assets. Five of these collaboration models were initially integrated in SARA (turn, layer, ownership, hierarchy-based and unconstrained examples) and the platform enables the definition of new ones. Thanks to the reusability of its components, during the development of an application, SARA enables focusing on the application logic while avoiding the implementation of the communication protocol, data model handling and orchestration between the different, possibly heterogeneous, devices involved in the collaboration (i.e., mobile or wearable AR devices using different operating systems). to describe how to build an application based on SARA, a prototype for HoloLens and iOS devices has been implemented. the prototype is a collaborative voxel-based game in which several players work real time together on a piece of land, adding or eliminating cubes in a collaborative manner to create buildings and landscapes. Turn-based and unconstrained collaboration models are applied to regulate the interaction. the development workflow for this case study shows how the architecture serves as a framework to support the deployment of collaborative AR services, enabling the reuse of collaboration model components, agnostically handling client technologies.

## 1. Introduction

Augmented Reality (AR) and Mixed Reality (MR) have spread rapidly to become familiar technologies. Thanks to the emergence of more powerful devices (both mobile and wearable ones) and evolved Software Development Kits (SDKs) and frameworks that facilitate development of AR and MR applications, its future seems brighter than ever. This is such that some sources indicate that, for example, 100 million customers will use AR both in physical stores and when making on-line purchases [1]. Much of the potential of AR and MR is shown in collaborative scenarios. Collaboration in AR and MR is not a new issue and quite a few systems have already been developed mainly to provide solutions for specific applications, thus lacking generalization capabilities.

This paper introduces SARA, a microservice-based architecture to facilitate the deployment of Shared-Augmented Reality experiences and Applications. SARA has three main strengths. First, it allows building AR-based collaborative applications regardless of the end-devices (mobile or wearable ones). SARA is also able to automatically orchestrate the rules for the collaboration policy needed for the target application. To do so, we defined a set of collaboration models which implement the most frequent collaboration strategies, such as turn-based, layer-based and hierarchy-based ones. Finally, the architecture facilitates the adaptation of non-collaborative applications to make them collaborative, and even to turn those non-AR applications into AR ones. SARA architecture is tested in this paper through a case-based methodology—technology empirical evaluation—to complete a proof-of-concept validation. The case under study is the development of a voxel-based collaborative game, which is implemented to demonstrate the development life-cycle and the management of unconstrained and turn-based collaboration models.

This paper is structured as follows. First, in Section 2 we provide an overview of the current state of the art regarding AR collaborative systems as well as collaborative Augmented Reality architectures. In Section 3 we present an analysis of the collaboration over AR, analyzing its main characteristics and functional features. To do so we divide the collaboration schema into three different layers (visualization, interaction and collaboration management) and we evaluate the workings of some examples of collaboration models such as turn or layer-based ones, among others. Next, in Section 4 we present the main characteristics of SARA, the architecture itself and its components. In Section 5 the data model is introduced, with the main entities handled within the architecture. Then, in Section 6, we present the process of building an application based on a specific and illustrative case. Finally, Section 7 concludes the work and summarizes future research lines.

## 2. State of the Art

In its most basic definition, Augmented Reality (AR) consists of superimposing digital content over a view of the real world. One of the most widespread interpretations today is the one from Azuma [2], who specifies that AR has to meet three basic requirements: (i) to combine the real world with virtual information, (ii) to be real-time interactive and (iii) to be presented in 3D. The concept of AR is closely related to Virtual Reality (VR) and even more to Mixed Reality (MR). In contrast to AR, in which the digital information is superimposed over the real world, in the case of VR the user is immersed within the digital world. Milgram and Kishino [3] introduced the *Virtuality Continuum* concept to establish a classification based on the user location: real/physical at one end and virtual at the other. The Mixed Reality term appears here as a mix between virtual content and information from the physical world, in such a way that digital content may interact with the real world. Presently, the *AR* term is sometimes used in substitution of MR by developers and users. For the purpose of this paper we will do this assimilation. Although we refer to the term AR we also assume that virtual content will be related to the physical world, thus being the proposals directly applicable to MR problems.

The concept of collaboration in Augmented Reality context took its first steps shortly after the first AR systems [4] emerged. One of the first approaches was the Transvision System from Rekimoto [5], which allowed multiple users to share virtual content graphics disposed in a table. This system also enabled the users to interact with the content by choosing and action (e.g., selection or manipulation) to be implemented through a physical device. Two more important concepts may be highlighted from this system. First, the ownership of the object has to be transferred among users. Hence, two or more users could not work over the same digital element at the same time. This concept of blocked ownership has been adopted in many of later systems, with its associated advantages and limitations (a user must wait for another user to end-up the interaction in order to do hers). Second, each user had its own 3D model database synchronized with the other ones by propagating modifications. The Studierstube system [6–8] substituted that process of sharing the same digital content between all the users by the concept of information layers. These layers could be shown or hidden to the users in order to control the information presented to them. The manipulation of the digital content and the interaction with it was performed through a physical device. Once more, the same concept of a master device and multiple slaves from Rekimoto appears here, since the master stores the current state of the virtual content, which is downloaded from the slaves side. Additionally, any changes performed to the state of the virtual content on the slaves side is propagated to the central state maintained by the master. A fact that makes the system interesting is that it can be seen more as a service provider for different applications such as education [9] and gaming [10] than a single application. One of the first systems that explored the idea of combining different platforms to perform the collaboration (mobile devices and stationary computers in this case) was the MARS project [11]. In this system, authors made a distinction between two different contexts where users might be, outdoors and indoors. An indoor user could manipulate virtual annotations of real world objects while the outdoor users would see them properly placed. Subsequently, both users could swap their roles. Despite the limitations imposed by the hardware required at the time of publication, this work stands out as one of the first platforms to integrate different types of devices in collaboration. With the VITA project [12] Benko et al. offered a system for off-site visualization of an archaeological excavation. They combined Head Mounted Displays (HMD), a tracked hand-held device, a high resolution display and a multi-touch projected table surface to enable the collaboration. From a high-level point of view, the collaboration was thought as the navigation, manipulation and browsing of digital data. Thus, each user had their own representation of the digital content and they did not share the visualization component. In particular, this system did not enable multiple users to manipulate the same digital element. Nilsson et al. [13] proposed the use of Augmented Reality to support collaboration between different organization in crisis management scenarios. In this case, AR content sharing between several users is explored, but adapting the visualization to each participant. The authors faced the process of sharing common organization-specific information on a map. The proposed solution consisted of decoupling the data (e.g., the position of a car on the map) from its representation. Hence, the associated AR component for a piece of information was generated based on the organization of the participant. In this system, each client device was notified when another user made any change in its own virtual representation (e.g., if a vehicle's position was changed), and the client internal AR representation was updated. Regarding the interaction, it was performed by means of a physical device (a joystick). Thus, only the user with that device was able to interact with the AR content. On the other hand, the DARCC system (Distributed AR for Collaborative Construction simulation) from Hammad et al. [14] offered a system for operating and locate digital construction elements (in particular cranes) in a collaborative manner using AR. The collaboration in this case was performed within the construction site, with all the participants visualizing the AR content. In particular, the digital content was aligned by combining GPS and device sensors to convey the idea of the assets being located in the physical world. Although the AR content was shared among all users, their interaction was limited to control a single associated element (the crane). Since the content alignment was done on GPS, the use of this system was limited to outdoor contexts. Apart from these works, there are many other examples of collaborative systems that can be divided according to their area of application: *Industrial* [15,16], *Architecture and Construction* [14,17], *Education and training* [18,19] and *Entertaining* [20,21]. An exhaustive review of collaborative systems based on AR and MR to date has been recently proposed by de Belen et al. [22].

The creation of frameworks and architectures to facilitate the development of new cross-platform applications was already raised previously in the literature, as previously stated. Some of these works focused on wearable devices [23] or smartwatches [24], infrastructure devices such as the Kinect [25] or the generation of adaptive web applications [26] and user interfaces [27]. Following the same approach of facilitating the development of cross-platform application but focusing on the Augmented Reality Context, Speicher et al. [28] developed the XD-AR framework, created for addressing the issues found when developing AR applications in a multi-platform context. More specifically, they focused on the more recent devices and frameworks used in that moment (e.g., Tango, ARCore and ARKit).

In industry context, Microsoft has presented their *Spatial Anchors* system [29], which is a cross-platform developer service to deliver multi-user Mixed Reality experiences. It enables storing and afterwards retrieving the position of the users' devices and their relation with the real world. Microsoft has also recently launched their *Spectator View* system [30]. Spectator View allows streaming the MR digital content from a HoloLens device (in the form of *holograms*) to be shared with other users who are carrying 3D hand held devices. Thanks to this feature, the digital content can be easily shared among several users. Spatial synchronization is performed among the devices to present the digital content aligned. Although it is a very powerful and easy to integrate tool, its main drawback is that the cross-platform support is offered by means of the Unity environment, which is itself cross-platform. When integrating it into an iOS or Android device, for example, it must be done through a Unity application, i.e., without using the native frameworks.

In summary, (i) there is not an agreement on what collaboration in AR implies, and although (ii) there are already proposals to facilitate the development of cross-platform applications in the context of Augmented and Mixed Reality, they are focused on a reduced and close set of devices. In general, these frameworks, tools or approaches do not take into account how to integrate new devices or different development ecosystems. Apart from this, the reviewed systems rely on limited collaboration models (e.g., by taking turns because there is only one physical control device that has to be passed) or present a decoupled collaboration (e.g., each user works independently without sharing the AR content [12]). Additionally, the collaboration model is conditioned by the implementation of the system itself. For example, some projects use the concept of shifts (either having to transfer a physical device [13] or by external regulation) while others handle the concept of ownership (so that participants can only interact with those AR elements that have been given access). Thus, since the collaboration model is limited by the implementation, it is difficult (if not impossible) to change that model to adapt it to other requirements.

The following section will analyze the characteristics of collaboration in AR environments, which will later be taken into account for the development of the architecture.

## 3. Analysis of the Collaboration in AR Context

*3.1. Definition of Collaboration in AR Context and Its Main Characteristics*

A possible meaning for "collaborate" is "to work jointly with others or together" [31]. Thus, from a high point of view, several users in a group may do some actions to achieve a certain goal. However, in addition to those common goals, it may be possible that each user has personal goals. Hence, *competition* can be seen as a sub-case of the collaboration. In this Section, we aim at analyzing the concept of collaboration in AR context to define a set of features or characteristics that have to be managed in an AR-oriented collaborative architecture. Three are the main characteristics of the collaboration:

**Number of participants in the collaboration**: First and most obvious, to perform a collaboration multiple users are required. Thus, the number of these participants may affect the design of collaborative AR systems.

**Individual vs. common service goals**: Depending on the application, there may be one or several objectives common to all users. However, it is also possible that each user has one or more personal objectives, which may not coincide with the common ones. As an example, let us suppose a multiplayer game based on AR. Players in a team may be willing to win, while each player individually may be targeting to improve his personal score. Another example could be a collaborative BIM-based (Building Information Modeling) tool in the architecture context. Several participants may work together to design a building, being each user focused on different components (e.g., one may arrange the electricity system while another may place the piping system). In general, each of them can work independently in their own copy of the building. However, in specific occasions both may have to work together and even reach agreements.

**Operational features to take into account when implementing collaboration**: to achieve the collaboration goal, some actions have to be performed by the users. Thus, there may be interaction with the AR content as well as communication between the participants. The implementation of courses of action in collaborative AR requires to determine aspects such as time and space management, interaction methods, visualization and, all in all, the collaboration model used to orchestrate the collaboration. we following comment on all these issues:

- *Real time vs. shift collaboration*: Time management restrictions are important when defining collaborative AR applications. It is possible for example that all the users are working with the AR content at the same time, which could be seen in *real time*. However, it is also possible that they work *in shifts* and that they do not even coincide in time. In this case, the state in which a user left the digital information must be able to be retrieved by a second user in order to continue working.
- *Shared physical space vs. remote*: If all participants of the AR collaboration experiences are located in the same physical space, it is usually said that they are co-located and they are working "locally". On the contrary, if the participants do not share the same space, it is usually said that they work "remotely". It is also possible to have a mixed situation in which some users are co-located and they are working remotely with another one.
- *Interaction methods*: According to the target AR platform, interaction with the virtual content is implemented in a different manner. For example, in hand-held devices such as smartphones, the main interaction method is touch-based. However, in case AR Head Mounted Displays (HMDs), usually the interaction is carried out by hand gestures and/or voice commands. Thus, the type of interaction will condition the design of the means to implement the course of action.
- *Information exchange channel*: in order to collaborate, exchange of information between the participants is required. This can be done by visual means, including annotations or check lists, speech or combined strategies.
- *Content alignment and profiled visualization*: Digital elements may have different views tailored to the specifications of different users. Thus, the visual aspect presented by AR assets may not be exactly the same for all users. Among these characteristics are the object position, scale and rotation as well as more advanced ones such as color or shape. For example, let us consider a 3D AR cube that is positioned on a real, physical table. If the AR content is *aligned*, the cube will be seen by all participants in the same exact position (setting aside possible drift errors). On the contrary, if the content is *not aligned* the same cube will be represented in different positions for each user. For example, one may see it on the table while another user may see it on the floor. However, both representations will refer to the same cube. In this case, being aligned refers not only to the position, but also to characteristics such as rotation and scale (e.g., it is possible that each uses sees the cube in the same position but with different sizes). On the other hand, the information displayed to one user may not be visible to another, or it may show added or removed elements. In this sense we say that the content has been adapted to the user's profile; in other words it has been *profiled*.
- *Collaboration model*: the collaboration model defines how users organize themselves to visualize and interact with AR content. This organization may come from themselves (they agree by speech), due to implemented control mechanisms (e.g., if the element of interaction has to be transferred) or by high-level rules (e.g., it is defined that they have to work in shifts).

*3.2. Conceptual Framework to Define Collaboration*

Based on the collaboration characteristics in Section 3.1, we next abstract three service layers where collaboration constructs are managed, from less to more enriched interactive scenarios: (i) the *visualization* of the Augmented Reality content, (ii) the *interaction* with the AR elements and (iii) the high-level orchestration of first-two components. Figure 1 shows an outline of the three layers and how they would be instantiated within an example application: the voxel-based collaborative game that will be later detailed in Section 6. Below, each layer is explained in detail.

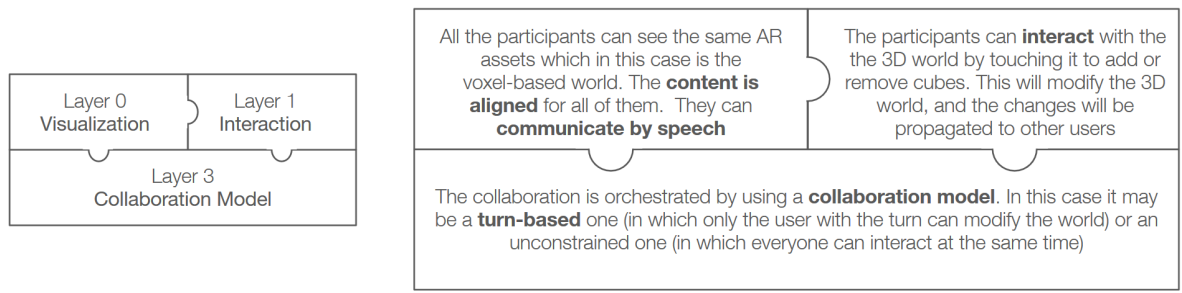

**Figure 1.** Collaboration Layers.

**Layer 0: Visualization.** This is the most basic level, since there can be no collaboration in AR if users cannot share AR digital elements. This level refers both *information sharing* and *alignment* of AR content. First of all, it is possible for all users to share exactly the same AR elements, i.e., that everyone can see exactly the same. However, it is also possible for the sharing process to be *filtered* according to some rules (e.g., based on hierarchical roles where the user who is at the top of the hierarchy has access to all the information levels while others can only see specific levels depending on the permissions based on their position in the hierarchy). Second, it is also possible that the *shared* AR elements present the same characteristics (e.g., position in the real world, rotation or scale) for all the users, or on the contrary that they are personalized for each one. At this point, the organization of the collaboration fully relies on the participants. Blocking and interaction have to be accorded by the users during their own collaboration. In this case, speech may be a useful way to regulate collaboration. In other words, the collaboration model is not restricted.

**Layer 1: Interaction.** Once the digital elements are shared among participants, the next step is to allow the users to *interact* with them. With that, we understand that the AR elements are able to detect user inputs and react accordingly. Let us suppose an AR cube that is positioned in front of the user, who is using a smartphone to visualize it. When the user *touches* the cube on the screen, it should be able to detect that touch and if necessary, e.g., to execute some application logic such as changing its color. It is also possible that instead of touching, the user performs a *drag* gesture, which could have a different associated logic, such as making the cube rotate on itself. Therefore, different ways of interacting with the AR elements can be present for a single device. Likewise, for each type of AR device, the ways of interaction with the AR content will be adapted to them. As was the case at level 0, the collaboration model does not present any restrictions. Users will have to organize themselves to regulate the interaction. Thus, the next layer emerges with the aim of automatically orchestrating the collaboration.

**Layer 2: Collaboration Management.** With layer 1, participants are given the ability to interact with AR content. It may be the case that several users decide to interact with the same element at the same time. In principle, there is nothing that prohibits or regulates this situation. How should a hypothetical AR cube react when one user moves it up and another user tells it to move down? in this case the logic of the application is wrapped with an external element, a *Collaboration Model* (CM). A Collaboration Model is compounded by a set of rules that control all the elements of the collaboration, which can be both visualization and interaction capabilities. The most basic Collaboration Model would be the *Unconstrained* one, in which no rules are specified. Thus, even in the case that the collaboration management is not explicitly defined, that situation can be associated with a Collaboration model. Another example may be the use of a *Turn-based* model. In this case, the interaction with the AR elements is restricted only to the user whose turn it is. In summary, by defining the rules that shape a CM, it is possible to control the flow of an application without modifying its basic logic. An important fact is that the same set of rules (i.e., the same Collaboration Model) can be applied to completely different applications. In other words, collaboration models are application independent and can be reused. Below we propose five types of high-level collaboration models to illustrate this concept.

1. *Turn-based model*: the interaction with the AR elements is managed by using a token. Only the user who has the token (therefore in her turn) can interact with the virtual objects. The interactions of other users with the AR content are in principle discarded.
2. *Ownership-based model*: in this collaboration model, each digital element has an associated owner. Thus, the visibility and interaction with each element may be restricted only to its owner. It is also possible that the elements are visible for all or a group of users with only the interaction being limited.
3. *Layer-based model*: This collaboration model is characterized by having one or many *layers*, to which the AR objects are linked to. One digital element may be present in several of those layers. Users on their side will have access to certain layers and they will be able to visualize and also interact with its associated digital elements.
4. *Hierarchy-based model*: in this case users are organized in a tree-based hierarchical structure. On one hand, the user at the top of the hierarchy has access to all AR elements, both for interaction and visualization. Changes performed by that user will always prevail over those initiated at a lower level of the hierarchy. On the other hand, users located in the lower branches of the tree will have limited both their visibility and their ability to interact. It is reasonable to assume that it will be the participants themselves who define these restrictions for their subordinates.
5. *Unconstrained model*: in this collaboration model there are no restrictions. All user interactions are processed and no control over the visibility of AR elements is established. In this case a FIFO (First In, First Out) policy will be applied.

It is important to note that these five models gathers the main and common collaboration schemes we identified and that there may be more possibilities, since the rules themselves are what define each of the models.

Which collaboration model to use entirely depends on the service goal of the application to be developed, i.e., SARA is not imposing any collaboration model but enabling choosing or combine the most adequate ones for the application to be developed. In Table 1 we illustrate how different collaboration models can be applied within two different applications. On the one hand, we propose the voxel-based world editor that will be described in more detail in Section 6. At this point, it is enough to indicate that multiple users need to be able to interact with a portion of land formed by cubes (voxels), so that by adding or removing those cubes they can create landscapes and buildings. The resources to be handled will be (1) the land itself, (2) the voxels (which may be added, removed or change their color) and all players will have identical permissions. On the other hand, we propose a prototype application for visualization and management of SmARt city resources (the SmARt City viewer). In its most basics, this prototype allows users (from the municipality, service providers, operators) to inspect injected data sources on a 3D representation of the terrain. The key concept of this prototype is that AR resources visualization will be based on the roles assigned to the users. Table 1 summarizes how each collaboration model may fit or not for both applications.

It is also important to note that an application could also integrate different collaboration models, e.g., depending on the users' role and the interaction level. For example, in the SmARt city scenario, the municipality could establish a *hierarchy-based* model over service layers to force and control specific aspects of the city management; at the same time, urban service operators working the same hierarchical level could own and control their network of resources (*ownership-based* model), while interacting with others providers on shared resources following a *turn-based* model. Figure 2 illustrates an example of the combination of multiple collaboration models.

**Table 1.** Use cases of the collaboration models and it suitability for the voxel-based game and the SmARt city prototypes.

| | Use Case | Voxel-Based Prototype | SmARt City |
|---|---|---|---|
| Unconstrained | When parties coordination is not needed or relies on human means | A choice if every player is empowered to add and remove cubes simultaneously and coordination relies on peer-to-peer communication | Not adequate, as it is assumed that the number of services service providers and resources in the city will be large enough to need at least information filtering services for visualization |
| Turn based | When resources need to be univocally controlled by a single user at a given moment in time | A choice if every player is empoweredto add and remove cubes and coordination needs to be orchestrated to avoid conflicts and organize interaction | A choice in case two or more service providers share a common resource and may have control over it: e.g., a signage panel in which alerts can be presented may be admitting new information in case the control token is free |
| Layer based | When resources need to be grouped and the access to those groups has to be managed | It does not apply, as this application only contains a single node: the terrain mesh that players will modify | Layers will be in this application associated to a given service (cleaning, garbage collection, air quality monitoring, traffic controlling…). Each layer will contain a set of nodes in it and will be secured by applying role-based access. Users authorized for each layer will be able to visualize the resources on it. |
| Ownership based | When a user owns one or more given nodes with full control over them. These resources could be within the same layer or not | Not applicable, as this application only requires a single resource for common interaction: the terrain mesh that players will modify | Applicable to determine e.g., which operators can access and visualize the state of specific resources. |
| Hierarchy based | When a hierarchy of roles is needed to control the different resources | The basic application is built on players with the same role, thus the hierarchy-based model does not fit. In case hierarchies are implemented, it is assumed that a user with upper role (game master) would have at least to approve the initiatives of the others | A hierarchical model does fit the collaboration control in this application, as there will be users with different roles that will be authorized to access functionalities depending on their role, e.g., municipality (overall control), service providers (overall monitoring of the service, interaction with shared resources), service managers (tactical decisions in a service layer and over owned resources) and operators (visualization and information generation over authorized resources) |

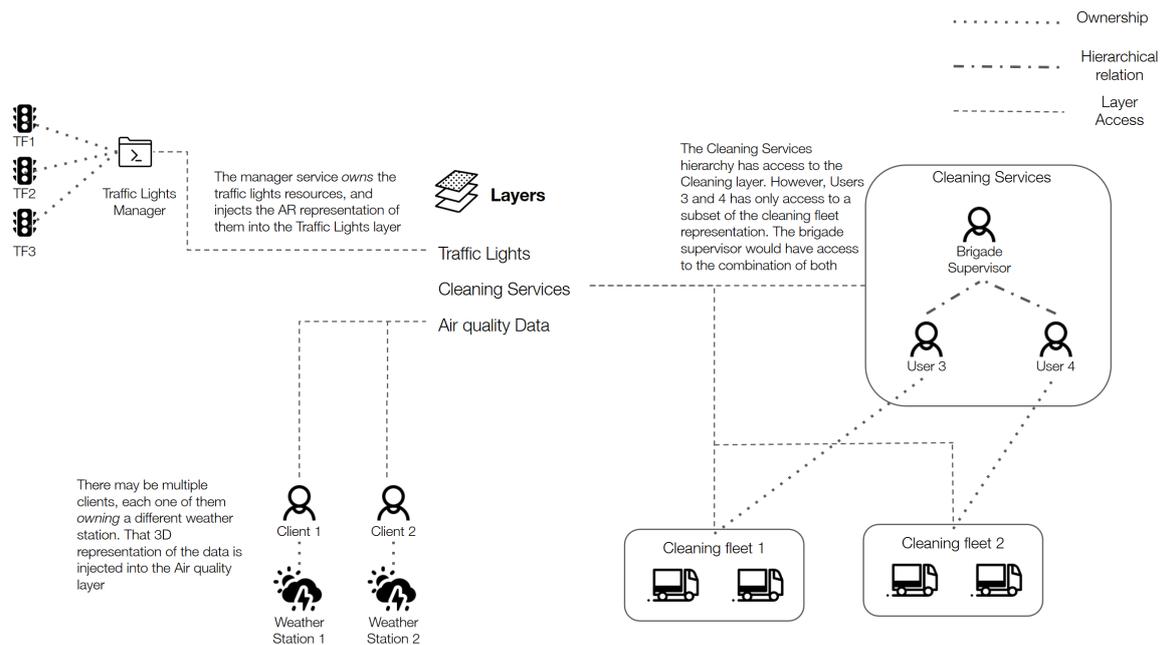

**Figure 2.** Collaboration Layers.

As can be seen in Figure 2, a *layer-based* model is present, with three managed layers: *Traffic Lights*, *Cleaning Services* and *Air Quality Data*. Access to the *Traffic Lights* layer is granted to a Traffic Lights Manager, which *owns* the nodes associated with the traffic lights. At the same time, access to the *Air Quality Data* layer is granted to multiple clients. Those clients generate and *own* the nodes associated with different weather stations. These clients could be institutions or citizens, being the ownership associated with them. Also, at the same time, a hierarchical-based model may be used to establish the control over the elements of the *Cleaning Services* layer. The members further down the hierarchy would have access to a subset of the vehicle fleet, while the supervisor would have access to the entire set. By combining multiple models of collaboration, conflicts could arise. For example, if a user had been given access to a layer with a certain element, which belongs to another user, should he be able to view it? Looking ahead, for future versions of SARA we propose the use of priority rules to manage, which should help in those advanced scenarios.

## 4. SARA: Main Characteristics and Architecture Components

The development of the architecture has been carried out based on the challenges detected in the solutions of the literature and the collaboration characteristics exposed in Section 3. SARA's name comes from "an architecture for **S**hared-**A**ugmented **R**eality experiences and **A**pplications" and its characteristics are presented below.

1. **Multi-user management**: Since SARA's main objective is to facilitate the development of collaborative AR applications, multi-user management is a vital concept of architecture. With this characteristic we refer to both user access control (e.g., logins, logouts) as well as the exchange of information between participants in the collaboration.
2. **Session management**: in order to orchestrate the collaboration, SARA uses the concept of *Session*. Several users may connect to the same Session, each session with a different functional concept and objective. The AR content associated with a Session may be shared among the participants in the collaboration. Each session will be treated independently within the system, so the same instance can be used for handling totally different applications at the same time.
3. **AR content visualization management**: SARA is able to manage in a flexible way the visualization of AR assets, i.e., it is capable of controlling features over the shared AR element (e.g., it is possible to control that visualization based on permissions) and also who visualizes the common features. This management can be carried out from the functional logic of a Session or externally through the use of collaboration models.

4. **Device-optimized interaction management**: in addition to display AR content, the system handles interaction with AR content adapting it to the specific standard interaction means of the target device. Each device has an inherent form of interaction associated with it. SARA integrates all these interaction types. As an example, when a user works with a smartphone the main interaction is performed through its touch screen. On the contrary, when the user is wearing an HMD such as the HoloLens, the main interaction is performed by hand gestures. However, the user expects that *a touch* on a digital object and *a tap gesture* to have a similar functional meaning. Moreover, interactions from different device types are translated to a common understanding.

5. **AR platform and development framework independence**: the whole architecture has been designed as a platform-independent environment. This feature means that different devices can cooperate over the platform. From smartphones to wearable AR devices, going through standard platforms such as computers, SARA provides a cross-platform information exchange system. Besides that, the creation of new end-point applications for the architecture is platform and framework independent, which allows developers to work with the tools of their preference.

6. **Location independence**: the collaboration between participants can be performed both *remotely* and *locally*. This process is transparent to the user, since all she has to do is connect to a session. Then, according to the session details, if required, the alignment process will begin.

7. **AR Content Alignment**: in case that AR content alignment is required, the system establishes it according to the requirements of the session. To do so, SARA offers *different strategies*. The basic mode of operation is intended to be through the use of markers, which are 2D physical images located on the real world. Although it would be ideal, the process of aligning cloud of points (also known as SLAM maps [32]) extracted from the real world reveals such difficulty that it is proposed to be integrated in a future version of the architecture.

8. **Time decoupling**: Collaboration is allowed to take place both in real time (or near-real time) with all the users collaborating at the same time and in shifts, with the participants not being coincident in time. To do so, SARA keeps a central state of the collaboration session must be maintained and changes on this state can be made at any time. Later, that state may be recovered on demand (e.g., if a participant connects later to the collaboration).

9. **Extensible**: It is possible to add new features to the system easily. This affects both the addition of new services with new functionality to the system, and the adaptation and extension of the associated data model.

10. **Independent of the communication network protocol**: in order to cover as many devices as possible, the system offers a transparent way of establishing the communication. At the moment of truth, there are certain frameworks and platforms that present different difficulties to work with certain protocols, hence the option of choosing is given to the developer. The use of TCP, UDP, Websockets and MQTT is currently included.

11. **Logic decoupled and reusable**: Thanks to SARA, all the logic of the application can be implemented for a single device/platform, which we call the *provider*. This *provider* will inject AR content to a session and the other participants will send interaction events to content. This second group is labeled as *consumers*. The main core of *consumer* clients is common to all the applications and hence it can be reused and taken as starting point for development. Over that core, some logic such as interface management may be implemented, but it is not strictly necessary.

12. **Adaptation of non-AR, non-collaborative applications**: by using SARA it is easy to inject the content of applications that were not generated with Augmented Reality in mind and expose it to users. It is also possible to update applications that were not initially developed as collaborative ones to allow that functionality.

Figure 3 provides an overall view of the architecture: a set of services will enable the communication among the participants and the management of the collaboration itself. Additionally, several SARA clients are depicted in the figure. Each of those clients will be used by the participants to collaborate, either using them one at a time or by combining several of them. Further details are given later in Section 4.2.

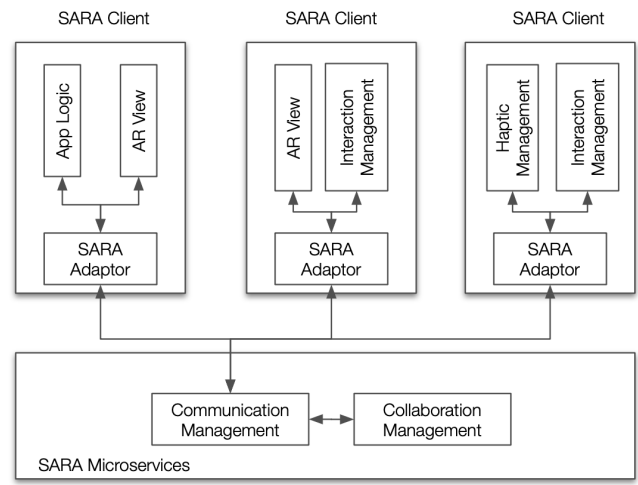

**Figure 3.** Overview of the SARA architecture, with its communication and collaboration management microservices and clients.

The architecture has been designed based on the microservice paradigm [33]. This decision has been taken to maintain each service loosely coupled and small in functional terms wanting to assure scalability and reusability. Thanks to this, in general the architecture will keep on working even if a given service is down (only it will not do it in case the service is managing communications, Section 4.1). Figure 4 shows the different services implementing the full architecture.

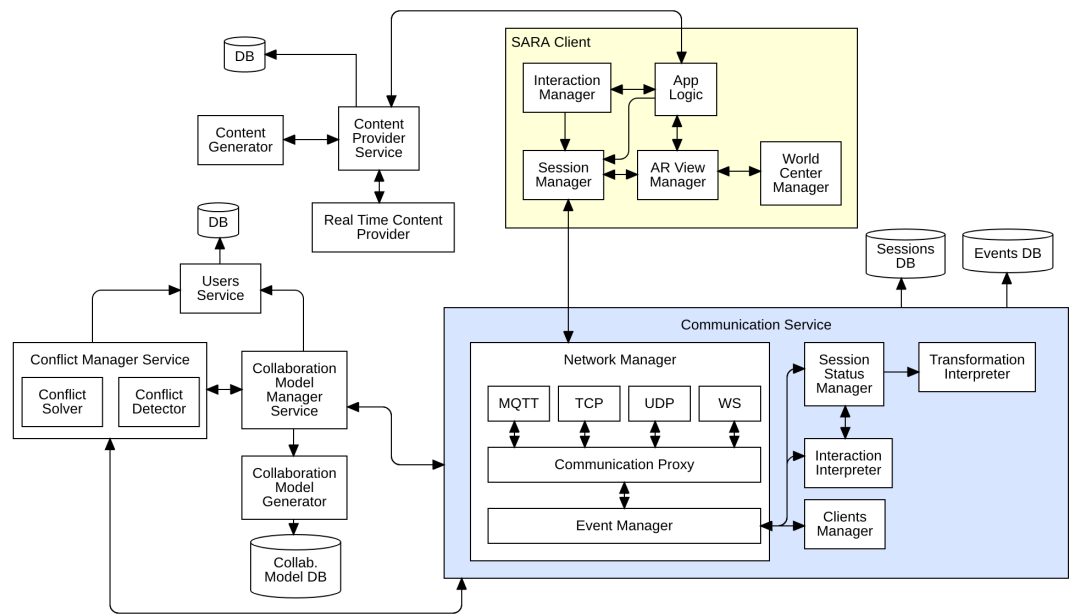

**Figure 4.** Service-oriented architecture diagram for SARA.

The architecture may be divided into three main building blocks: (a) the *Communication Service*, (b) the *SARA Client* and its associated services and (c) the *Conflict Manager and Collaboration* services, to be explained below.

*4.1. The Communication Service (CS)*

This service is the most critical of the architecture since it is in charge of managing the communication among all the others; i.e., in case this service is not running, the rest of the system will not work. Figure 5 illustrates an example in which three users participate in the collaboration.

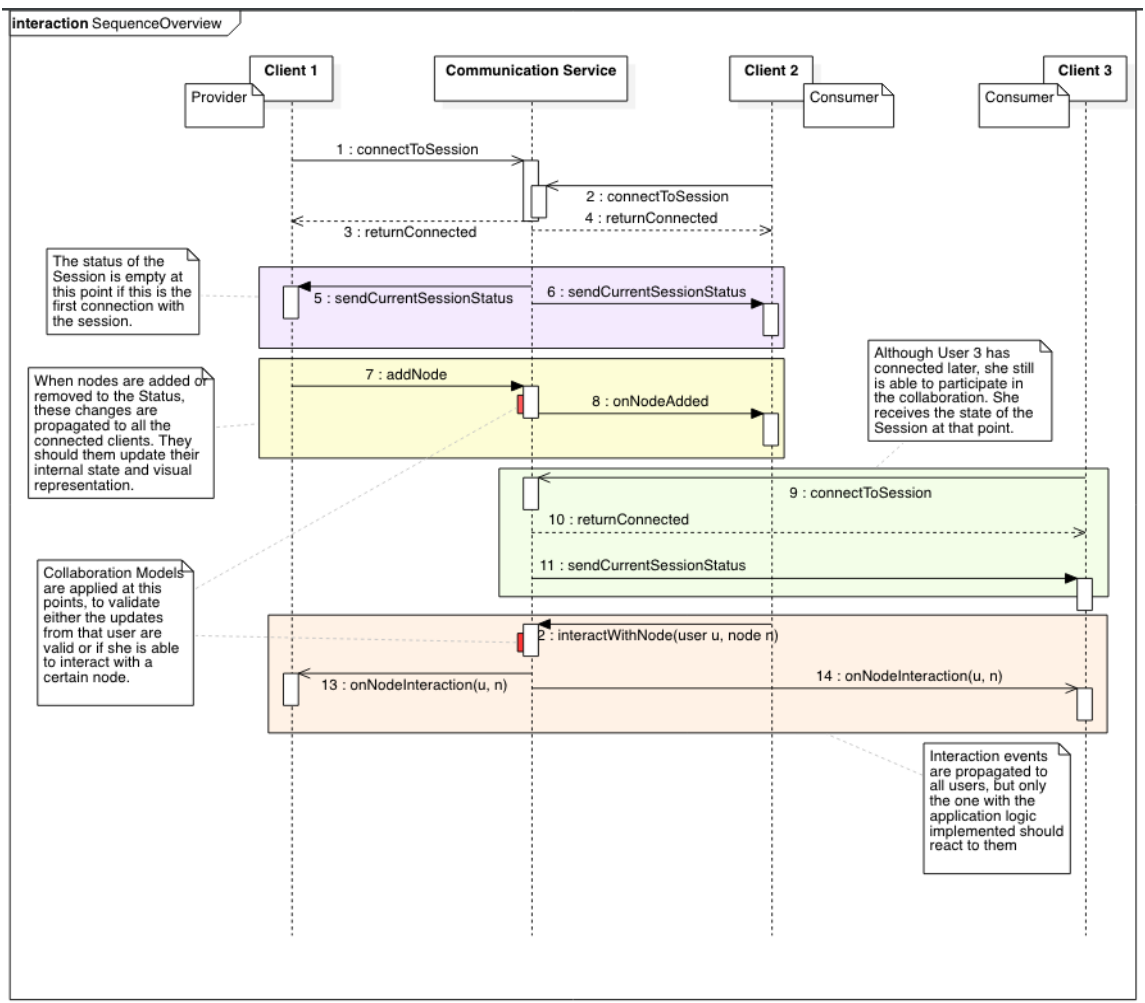

**Figure 5.** Sequence diagram that illustrates the communication between users and the *Communication Service*.

Inside the *CS* there is the *Network Manager* component, which is the main entry point to this service. When a *SARA Client* or some other services require to communicate with the *CS* they must establish a connection by using one of the possible technologies this service offers. In this case the possibilities are MQTT, TCP, UDP and WebSockets, which have support in the vast majority of development environments and programming languages. Once that connection has been made, the *CS* knows which kind of connection was used and the same method will be used for the communication. All the messages/events received are abstracted from the communication layer in the *Communication Proxy*. From there, all events can be understood by the *Event Manager*, which will parse them and react to their content.

There is also a component labeled as *Clients Manager*. It is in charge of storing the information of the connection with the clients and to storing their information. User credentials are not stored in this component, but only connection information.

The *Session Status Manager* stores the information of the sessions handled by the system. It reacts to update events of session status and apply them to the proper *Session*. Since different development platforms use different coordinate systems to arrange the digital 3D AR content, it is mandatory to have a component in charge of establishing a common framework. Hence, the *Session Status Manager* can cooperate with the *Transformation Interpreter*, with the aim of maintaining a consistent status of the *Session* understandable by all. The most common operation to be performed will be the Z-axis inversion of the coordinate systems, because some systems use a right-handed coordinate system and others use a left-handed one.

Finally, the *Interaction Interpreter* reacts to interaction events by translating them to a common representation. Once that is done, those events can be applied to the required *Session*. For example, it is possible for a user to interact with an AR element by touching a smartphone screen while another user does so by gestures while she is wearing an HMD.

*4.2. The SARA Client (SC)*

It represents the combination of a user and a device. The collaboration is established among different instances of these Clients. Please note that each Client may have a different final implementation, i.e., the main components must be adapted to the target client devices.

The entry point of the *SC* is the *Session Manager* component. It offers a standard way to communicate with the *Communication Service* in terms of network managing and events handling.

The *AR View Manager* receives *Session* events and is in charge of updating the AR elements that the user sees.

The *World Center Manager* must be implemented in such a way that a certain point of the real world is set to be the origin of the coordinate system associated with the *Session*. It is possible to implement manual policies (e.g., to allow the user selecting a certain point) or automatic (e.g., to use marker detection).

The *App Logic* represents the specific logic of the application embedded in this client.

Finally, the *Interaction Manager* component must fit the type of device in use. For example, in smartphones or tablets it must react to screen touches and to detect if some AR component was touched. If so, it asks the *Session Manager* to send the proper event. In other devices such as HMDs for example, those events can be generated based on hand gestures, user positions or voice commands for example.

With the idea of increasing the capabilities of SARA Clients, they may have access to *Content Provider Services*, which offer contents that can be integrated in the *Session* status. From there, it can be shared with other clients that may have no direct connection with the content providers. Their functionality can be offered in any form, as long as clients are able to access it. Finally, *Real Time Content Providers* and different *Content Generators* can be exposed to the system through the *Content Provider Services*. we can imagine a scenario in which, for example, real-time external information is integrated into the session's status as Internet of Things information or social media data elements.

*4.3. The Collaboration Services*

Collaboration Services comprise several services that allow to alter different characteristics of collaboration between users.

The first service is the *Collaboration Model Manager (CMM)*. It stores information about the collaboration model associated with a certain *Session*. Each collaboration model has a set of associated rules. When the *Communication Service* receives *any* event, it is validated using the *CMM*. This validation is done by comparing the event against all the rules associated with the collaboration model. If the event goes through all of them, it is valid, so the content of the event may be applied. Otherwise, by default it is discarded but other policies could be implemented too.

The *Conflict Manager Service* has two main components, the *Conflict'Detector* and the *Conflict Solver*. The first is in charge of detecting conflicts (such as two users applying updates concurrently or conflicts generated when applying the *CMM*). Once detected, the *Conflict Solver* reacts to these conflicts and manages responses. For example, in the concurrent updates scenario, it can merge both events into a single one and send it to the *Communication Manager* as solved.

As the last component, the *Users Service* offers functionality related with the user's data. It also has access to its own database and stores information such as hierarchies of users, which can be queried by the *CMM*. Additionally, this service offers functionality such as registers and removal of users, as well as logins and logouts.

Thus, SARA wraps the minimum functionality required to enable the development and deployment of collaborative AR applications. Figure 6 illustrates the difference between the development of collaborative AR applications without using SARA and using it.

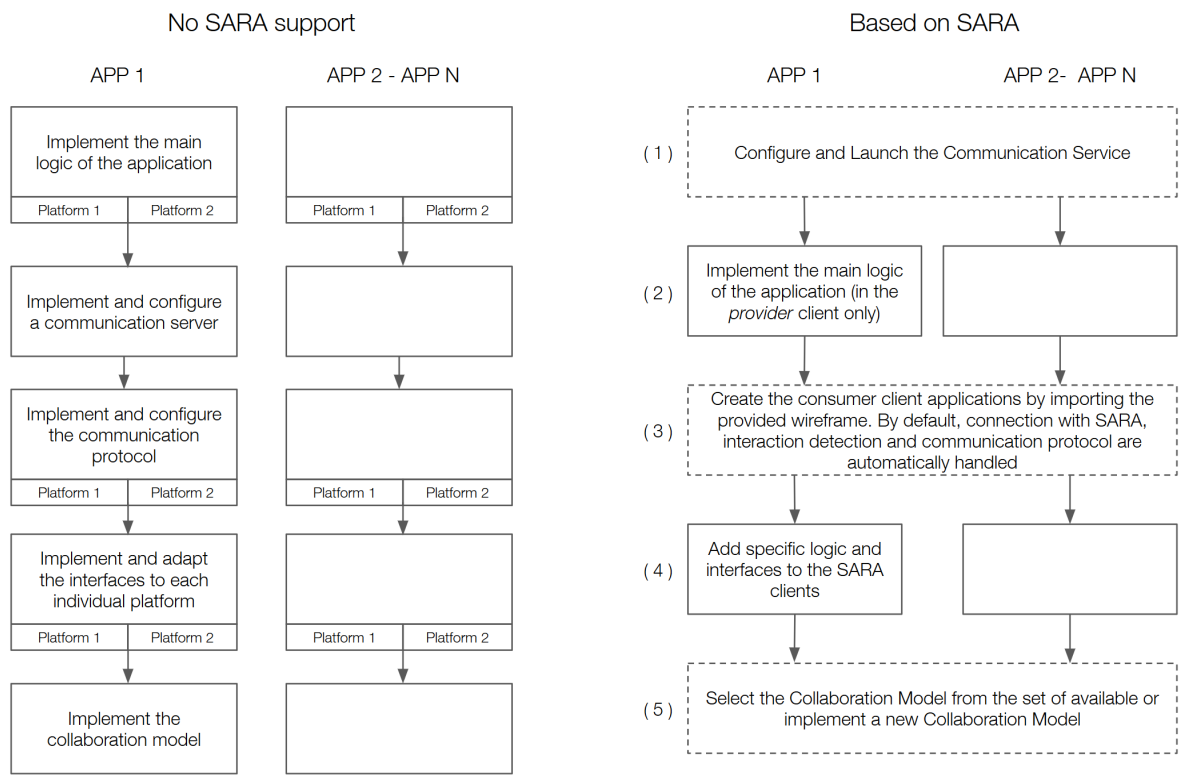

**Figure 6.** Sequence diagram that illustrates the communication between users and the *Communication Service*.

As Figure 6 shows, when implementing collaborative AR applications without using SARA, the development of each new application is done independently of the others. Thus, the application logic should first be implemented for each of the platforms on top of which the application needs to run. For example, in case the objective is to create an application that initially provides support for HoloLens and iOS devices (i.e., Platform 1 and Platform 2), it is required to implement the logic in a dedicated application for each of them. If later it is needed to give support for Android systems, that same logic should be implemented for Android. Once the logic has been implemented, a communication server that allows the exchange of information between clients has to be implemented and configured. Some previous server structure could be reused here, but it should always be adapted. Once done, the message exchange protocol between the clients and the server needs to be defined and implemented, and this had to be done for all the platforms involved. Next, the graphical interfaces and interaction methods of each platform have to be created and tailored. Finally, the collaboration model needs to be be implemented from scratch, possibly within the application's own logic, which means that it cannot be easily reused. By using SARA, there are three blocks in which development is accelerated. First, SARA provides a *Communication Service*, which does not require any implementation and can be directly launched (with some minor configurations such as port information). Thus, it will accept client connections and handle the events broadcasting. Second, when developing the platform clients, SARA already provides an application skeleton that once imported, establishes the connection with the *Communication Service* and receives/sends the session status. At this point it would only be necessary to create the interfaces adapted to the application. Third, SARA offers the possibility to choose from a group of already implemented collaboration models that can be applied independently of the application's functionality (although depending on it, some models will make more sense than others). In conclusion, SARA supports the development of collaborative AR applications by agnostically managing the AR content display device. SARA is built under the view of reusing a set of conceptual collaboration models, thus these assets are available for customization and integration.

Once the main components of SARA has been presented, in the next Section the Data model in which SARA is based will be presented.

## 5. SARA Data Model

One of the objectives of the SARA architecture is to be platform and framework independent. With this goal in mind, we have established a minimum data model that can be easily adapted to the most common concepts used by the main platforms and development frameworks. Most of them (e.g., Unity, ARKit and ARCore), work with the concept of *Scene*. This *Scene* groups in a hierarchical way a series of elements such as lights, cameras, and 3D models. In this case we have sought that the translation between these elements and those of our architecture to be as direct as possible. Figure 7 gathers in a class diagram the main entities of the model. In all the class diagrams presented, the prefix "SARA" has been added to those classes that could be confused with those of the development frameworks. Next, each entity will be presented in detail.

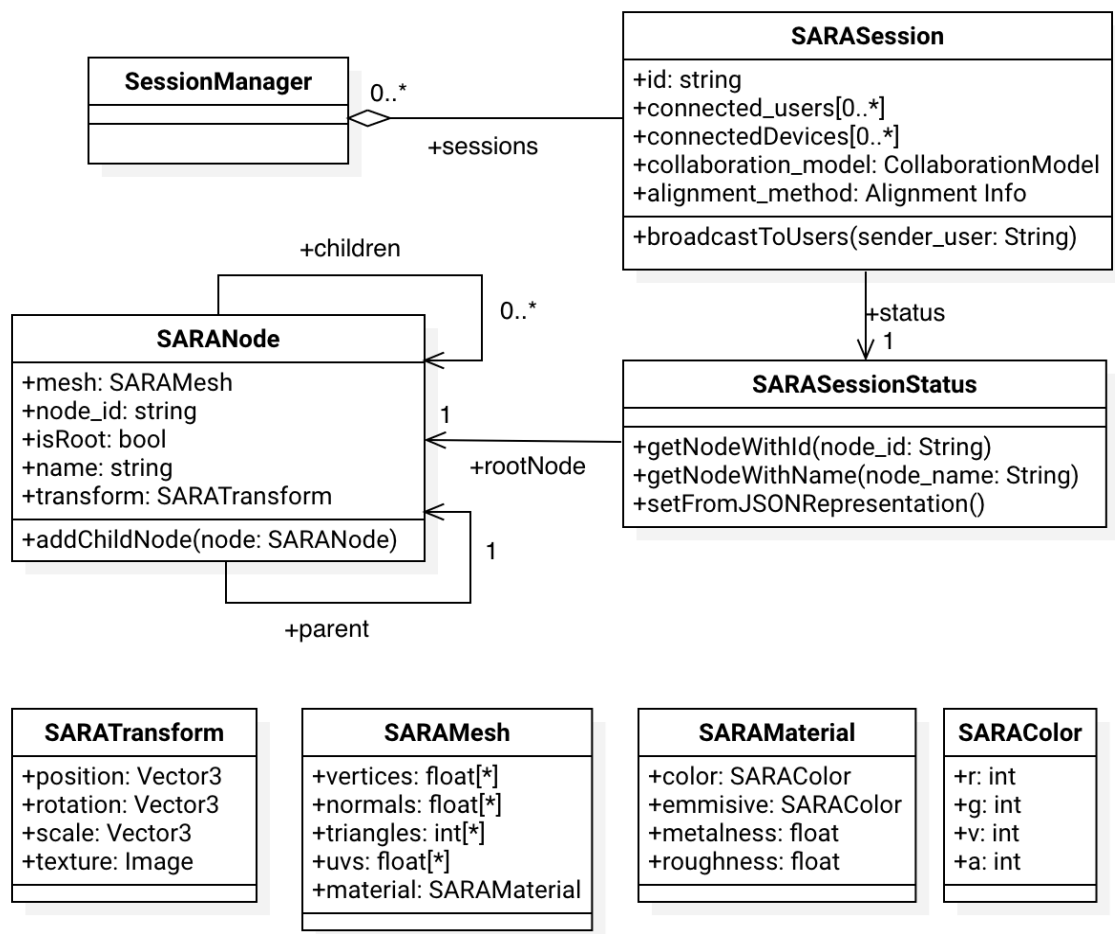

**Figure 7.** Class diagram of the most basic data model handled in the architecture.

*5.1. Sessions*

The main concept of the data model is the *Session* which can be translated to the concept of a *Scene* as previously mentioned. However, there is a slight change in the interpretation of the concept. Since we want to provide collaboration capabilities, we have to enable several users to share AR content and interactions. Hence, multiple users may connect to the same *Session*. The digital elements associated with a *Session* will be shared among the connected users. SARA is capable of handling several *Sessions* at the same time, each with its own resources, its connected users and its functional objective.

Each *Session* has a *SessionStatus* associated, which includes a reference to a root *Node*. This *Node* also finds its counterpart in the aforementioned platforms (i.e., a GameObject in Unity and a SCNNode in ARKit for example). A *Node* keeps a reference to its parent and its children, which make up a tree-like hierarchical structure referenced by the root node. Besides that, a *Node* may or may not have an associated *Mesh*, which stores the graphic information of the 3D model. It is possible to have *empty Nodes*, that only have a *Transform* (a position, rotation or scale) but no associated *Mesh*, with the aim of grouping several nodes under the same parent.

Regarding the *Mesh*, we have made the decision to keep it as simple as possible. To do so, we only store a list of vertices (each three numbers on the array form a vertex), a list of triangles (each number references the index of a vertex of the vertices array while every three indexes form a triangle) and a list of normals (each one associated with the vertices). we have decided to define the faces of the mesh as triangles to simplify the development process. However, if it is required to use faces formed by more than three vertices, it is easy to make that change.

Finally, each *Session* has an *Alignment Info* field associated. As shown in Figure 8, this class represents a common point from where different strategies can be implemented. For example, it is possible to set the *Alignment Info* to be *Marker Based*. This indicates that users must physically locate the marker image on his environment and once it has been found, the new coordinate system that encompasses the *Session* will be centered on it. In the future we are planning to allow the alignment to be established through the use of SLAM maps [32]. However, at this moment this process is not straightforward, since each platform and framework generates a different representation of the same physical space and they use different formats to handle that information. Today, some industry solutions are beginning to appear to solve this challenge, like for example, the Azure Spatial Anchors [29] solution from Microsoft. Since that process requires machine learning capabilities and it goes beyond the scope of this document, it has been decided to support it within the data model, leaving it for a future implementation. Additionally, the use of geo-positioning techniques like GPS may be integrated into the alignment in future versions. To end with the *Alignment Info*, there is a simpler possibility: to not align the virtual content. In this case each *SARA Client* should offer the user some kind of functionality to choose a real-world physical point that represents the previously mentioned coordinate system.

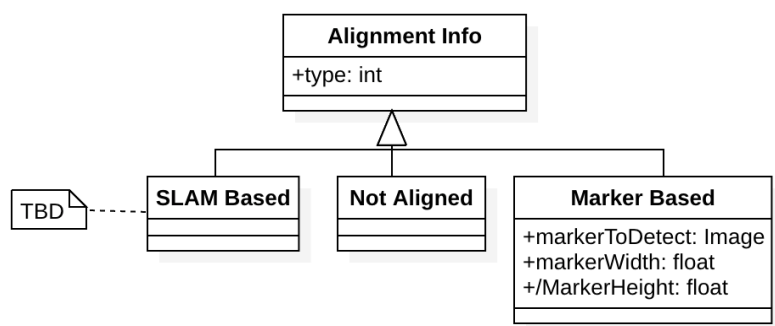

**Figure 8.** Detail on the Alignment Method associated with a Session.

*5.2. Events*

So far it has been presented how to encode the information that forms the *Sessions*. Now, it is time to expose the data model that allows the exchange of information between the different services of the architecture. The main entity used for this is an *Event* and its details can be found in Figure 9.

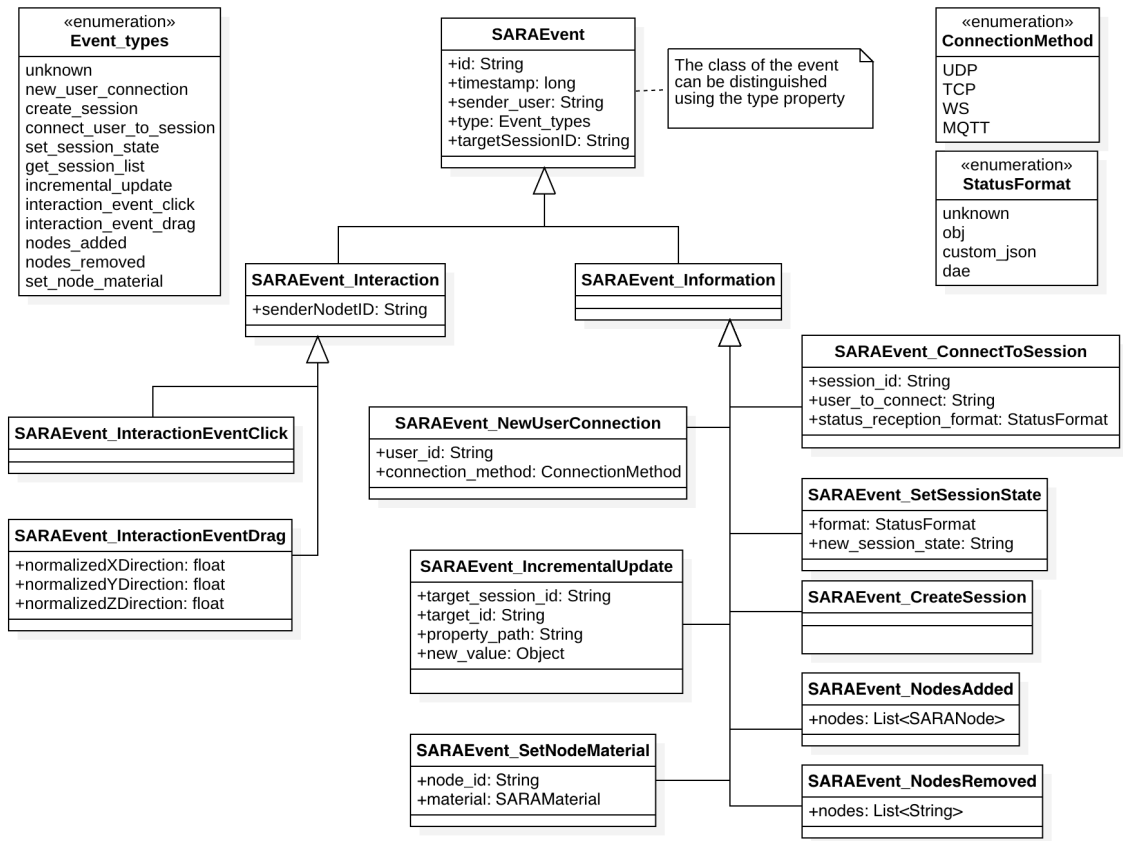

**Figure 9.** Data model of the Events handled by the system.

As can be seen in Figure 9 two fields are known for each *Event*: the user who launched it (i.e., the client id) and the target session (the *Session* to which the sender user is connected to). The type property is used to easily distinguish the type of event when decoding it. For the moment, there are two different types of *Events*: *Interaction* events and *Information* events.

On one hand, the *interaction* group encapsulates actions performed on a digital element, such as clicks (which would be a common interpretation of taps and touches) or drag. On the other hand, the *information* group includes events that update the information of the session or the connections, hence the name.

- When a client application wants to connect to the system, a *NewUserConnection* event must be generated. In this event the user that will be connected must be specified, as well as what we call a *Connection Method*. As will be explained later, the system permit the network connection to be performed using different protocols such as TCP, UDP, WebSockets and MQTT. Once the client has connected with the system, it can be easily known which of these protocols was used to establish the connection. However, in order to keep the data model clear, it has been decided to indicate this information within the event itself. At this moment, her connection information has been stored, but has not been connected to any *Session* yet. In order to do this, a *ConnectToSession* event is required.

- A *ConnectToSession* event must contain the identifier of the session to which the user is going to connect, the user itself and a *status reception format*. The reason of the last field is closely related to another of the types of information events, the *SetSessionState*.
- the *SetSessionState* event is used to completely replace the state of a session or to start it for the first time. As it can be seen, it has a *format* field and a *new session state* one. In order to maximize compatibility with as many platforms as possible, the system allows coding the state of a session using different formats. Some of them, such as OBJ or COLLADA, are the most popular on all platforms. However, the use of a custom JSON format is introduced here with the aim of allowing greater clarity to the exchange of events as well as greater flexibility. The idea is to export the status of the sessions (i.e., the nodes structure) to one of these formats and to encode the result as a base64 string. With this differentiation, different users may receive the same *SessionStatus* by means of different formats. It can be argued that this type of communication is not the most efficient. In future SARA implementations, the entire data model could be encoded as flat bytes, which would reduce the amount of information moved within the system (since the JSON format introduces redundant characters) and the communication times.
- Finally, as opposed to the *SetSessionState* which replaces the old state of a session with a new one, the *IncrementalUpdate* event is used to update specific properties of the nodes such as the position or the rotation. To do so, the id of the target node is required, as well as the path to the property. Let's take as an example an update event for the position of a node (which would be thrown, for example, when the node gets moved due to some kind of logic). The *property_path* would have a value of "*trans f orm.position*" and the *new_value* would store the new state of that property (e.g., an array like [1.0, 0.0, 0.0]). By using this event, selective changes can be made without requiring to override the whole state of the *Session*.

*5.3. Managing Collaboration Models*

The third pillar of the data model is related to collaboration models. In Figure 10 a class diagram of the collaboration models can be seen.

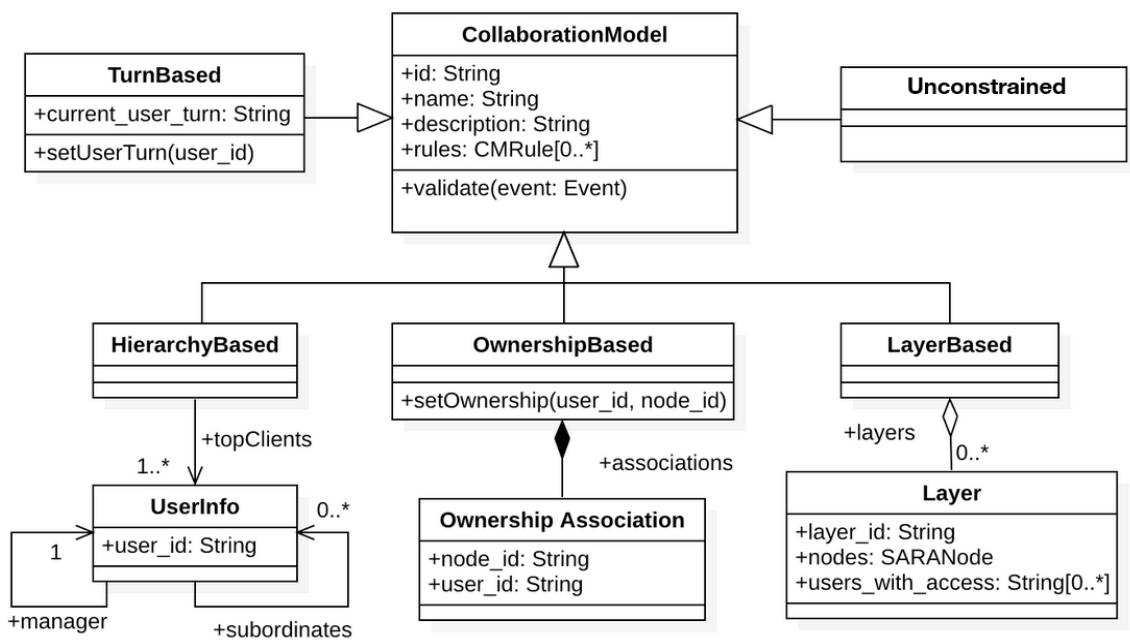

**Figure 10.** Data model of the Events handled by the system.

Five different types of models are proposed in this document. However, it is perfectly feasible to define new collaboration models by extending the basic ones. At the end, any kind of collaboration model has to implement the *validate* event through some kind of logic. This process can be done either by directly implementing the logic in code or by using a rule-based external inference system. Finally, in order to facilitate the understanding of the collaboration models we made five different models with non-overlapping logic. However, nevertheless, it is possible to combine multiple collaboration models at the same time. One example of this combination may be a game in which users can take turns actions, and in these turns they can only interact with those elements in to layers to which they have access. Thus, two models may be cooperating: a turn-based one for shift handling and a ownership-based one to restrict the access to the AR elements.

To manage the collaboration models, a special set of events is required. Figure 11 shows a possible set of those events for each collaboration model with the exception of the *Unconstrained-model*, which does not require events. These events can be used by developers or they can define their own set of events to define other types of collaboration models.

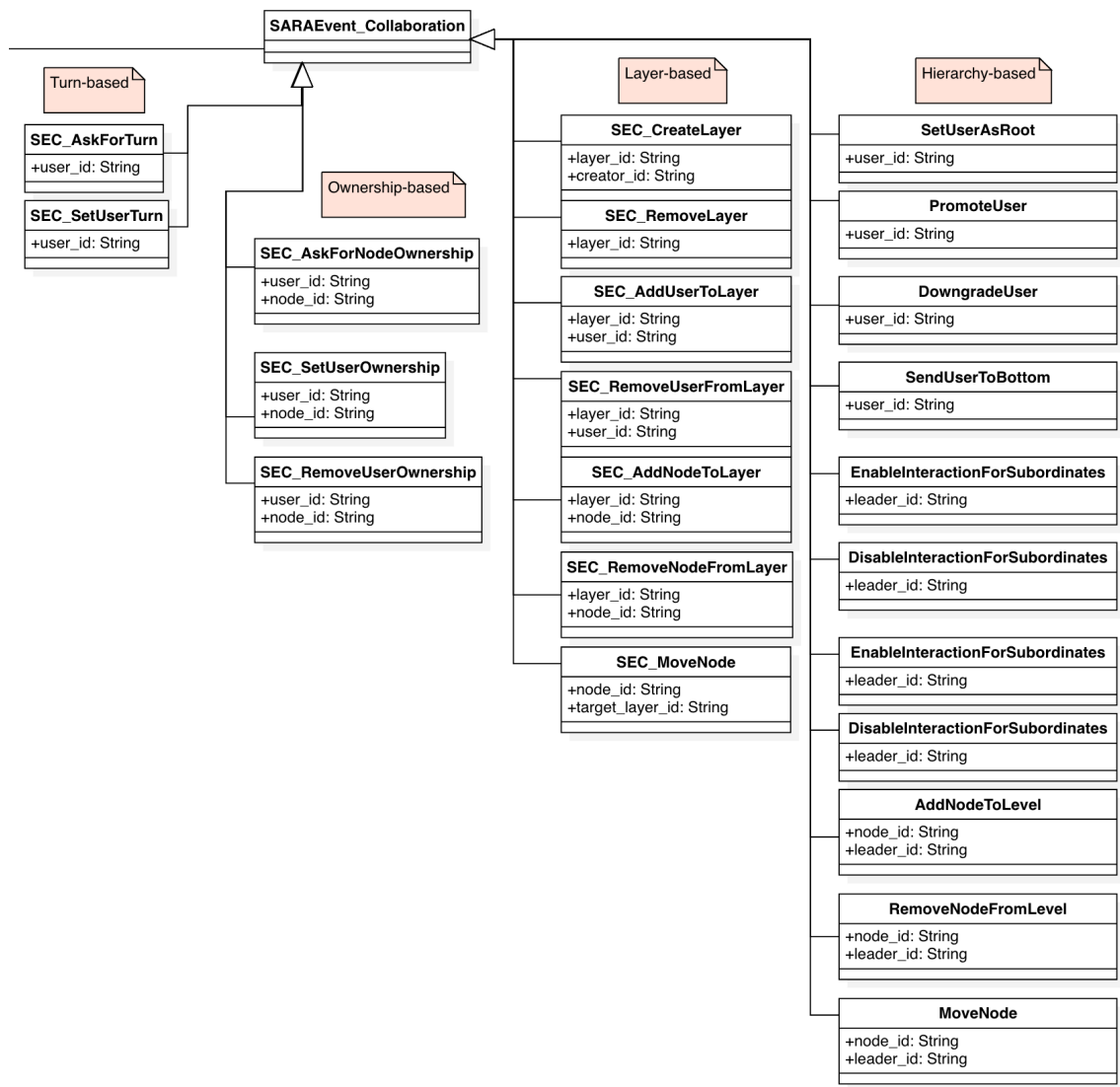

**Figure 11.** Data model of the Events required for each Collaboration Model.

## 6. Creating and Deploying Applications Over SARA: The Case of a Collaborative Voxel-Based Game

This Section presents an overview of the application development process based on SARA by using a case approach. The chosen case is a collaborative real-time voxel-based construction game, in which several users equipped with wearable and mobile devices may collaborate locally to achieve a common goal, by following different collaboration models (*unconstrained* and *turn-based* ones). The development case thus features specific design requirements that enables showing how SARA concept may provide to the development workflow. This prototype was inspired by Minecraft [34] and its more recent one implementation, Minecraft Earth [35]. In the original game, the player controlled a digital avatar inside a cube-based (voxel) world. Then, she can add or remove these cubes for creating landscapes and structures. In the Minecraft Earth implementation, the voxel world is constrained to a small surface and it can be viewed by means of Augmented Reality. Thus, in our prototype several players will work together on a piece of land, to which they can add or eliminate cubes in a collaborative manner. By default, a simple landscape appears, as a starting point for the users. From there, they can use three simple tools to manipulate the aspect of the world. These three tools are: (i) a shovel used to remove the cubes, (ii) a brush used to change the type of one cube and (iii) a block adder, which will add a block in the desired point.

This small prototype has a self-contained functionality that allows all the SARA capabilities to be easily applied. All the logic of the application, has been implemented using the Unity environment and initially it was only focused to be for a single user, allowing the participant to click on the screen to interact with the world. Thanks to SARA, AR collaboration capabilities can be easily added. To illustrate the developing process, we propose the combination of HoloLens (for wearable AR) and two iOS devices (an iPhone and an iPad, representative of hand-held ones). Table 2 gathers the main features of this prototype.

**Table 2.** Features of the voxel-world game prototype.

| | |
|---|---|
| Common Objective | To have fun while creating structures and landscapes together |
| Individual Objectives | Not specified. Users may distribute the work while playing |
| Number of Users | Unlimited |
| Location of the Collaboration | Mainly local |
| Involved Devices | Computer, HoloLens and iOS devices |
| Collaboration Model | Unconstrained model, users will have to agree the actions. This is done to mimic the operation model of the original Minecraft. Turn-based model also makes sense in this context |
| Temporality | Real time |
| Interaction Capabilities | All users have the same capabilities: selection of the tool to be used and selection of a point in the voxel world to apply the tool action |
| AR Content Alignment | The content alignment is done through 2D, physical marks |

The development process can be summarized in 5 steps: (1) configure and launch the Communication Service, (2) implement the main logic of the application, (3) create the SARA clients, (4) add specific logic to those clients and (5) selecting and applying the Collaboration Model. Figure 12 gives an overview of this process identifying the main tasks in each of the steps. Next, each of those five steps will be presented in more detail.

1. **Communication service configuration:** the first step is to launch the artifact that represents the *Communication Service*. In this case it consists of a Node.js application that has to run on a computer. This is one of the artifacts that have already been generated when designing SARA and therefore do not require implementation.

2. **Implementing the main logic of the application:** the second step is to create the application logic which, as has been said earlier, it will be a voxel-based creative game. It is also possible to reuse an application already implemented, in order to adapt it to collaborative AR. All this logic will be grouped within a *SARA Client*, which also incorporates the *Session Manager* component used for establishing the communication. To implement this point, the first step is to import the SARA library, which has already been generated and do not require intervention. Once done that, it is the turn to set some basic information within the SARA adaptor instance, such as the IP of the *Communication Service*, the Session to connect to or the format to be used for the session updates. When the application runs, the SARA instance will generate the required events to connect this client to the selected session. At this point, this client (the *provider*) is able to send events to the *CS*. It is labeled as *provider* because it will inject the AR content to the session. Once launched, it will make a general scan of all the elements within the Unity Scene, in this case the voxel world. From there, the updates of the session nodes would be made when some element in the scene changes. Additions and removals of cubes in the world will also generate update events, so all participants will receive the new state of the world. Figure 13 shows a screenshot of the Unity application with a small terrain already generated.

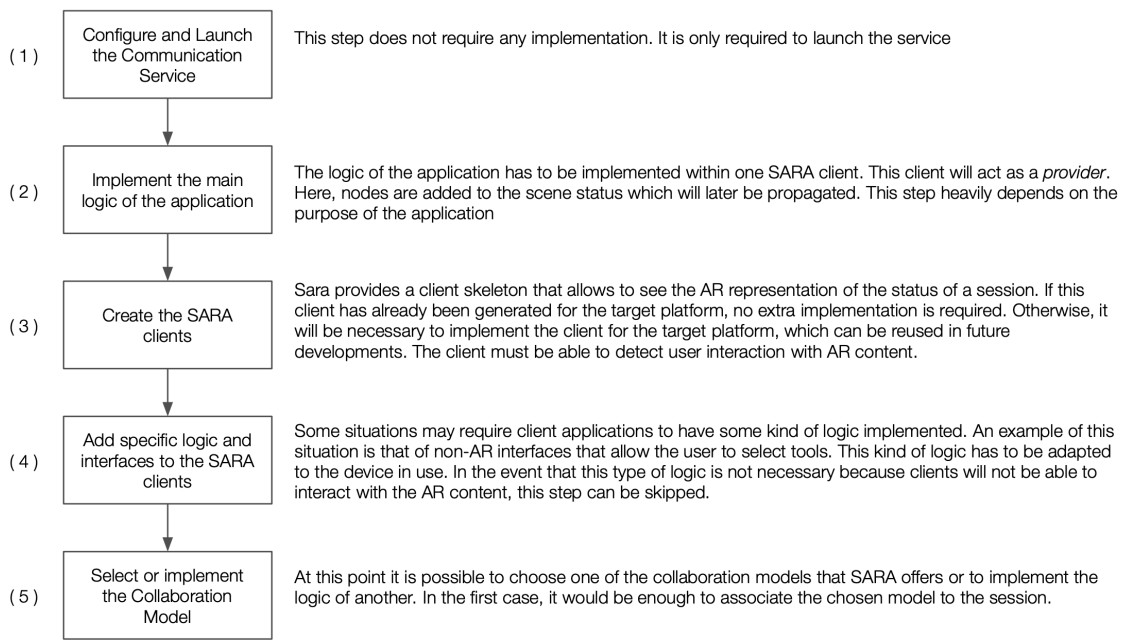

**Figure 12.** Overview of the five steps in which the process of developing with SARA is divided.

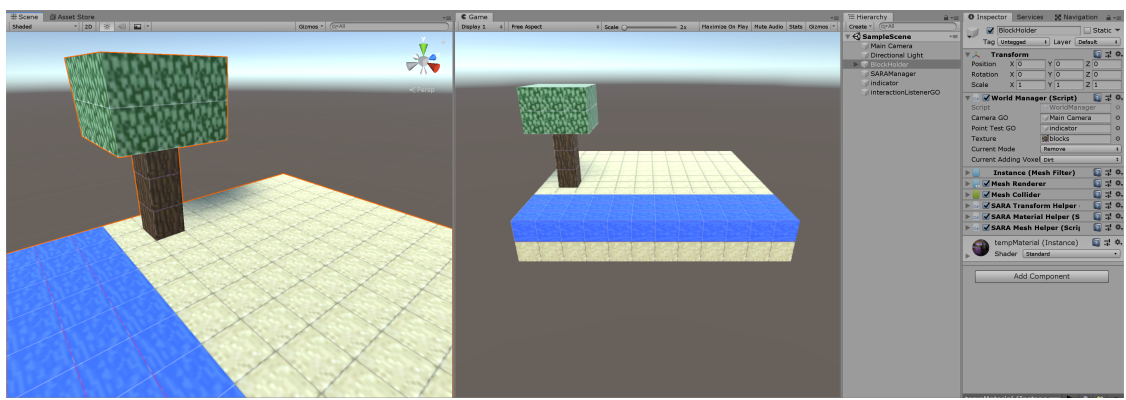

**Figure 13.** Screenshot of the Unity implementation.

3. **Creating the *SARA Clients* and enabling user interaction:** for the third step it is required to create the clients that will be used by the other participants, the *consumers*. To simplify, in this case we are going to stick to only two devices; a HoloLens device and an iOS one. Through these clients, users can see the session representation by means of AR. The implementation of this point is similar to that of the previous one. On either platform (Unity for the HoloLens and ARKit for the iOS one), the first step for generating the application is to import the SARA artifact. After setting the required parameters, the application is ready to receive the update events and to generate the proper AR representation. In both cases it is necessary to establish some way to detect user interaction with AR elements. One possibility is to detect user gestures (hand gestures in HoloLens and touches on the screen in the iOS client) at any time and if she is pointing to an AR element, then generating an *Interaction Event*. Another possibility is to add listeners to the AR elements themselves (GameObjects in Unity and SCNNodes in ARKit) with only the elements with these listeners generating those events. Finally, the functionality described up to this point is common to all SARA clients, i.e., it is independent of the application and could be used for any context.

4. **Adding specific logic to *SARA Clients*:** the fourth step is to implement the specific logic of the application for the users' clients. This point will be completely dependent on the application. In the example of voxel-based game, it is necessary to offer the users an interface from which they can select tools to use them. For example, in the case of the iOS client, a toolbar can be exposed to the users from where the action to be used is selected. After that, a cube is added at the point of the AR world that the user has touched. In the HoloLens case, the choice of the tool must be adapted to the form of interaction of that platform (e.g., voice commands, 3D tool palettes or 2D overlays may be used). Figure 14 shows a screenshot of the iOS client. From the tools in the bottom-left corner of the figure the user may select the tool to be used. Figure 15 on the other hand, shows a capture taken from the HoloLens, which in this case only receives the status of the session and has no interaction capabilities. Thus, the HoloLens participant can only inspect the AR content. we can imagine a scenario in which the playing field of some e-sport (e.g., League of Legends [36]) is injected into SARA. The spectators could then see an AR copy of that land where they wanted, but without the ability to interact. However, specialists could have access to a set of tools that allow adding annotations on the map.

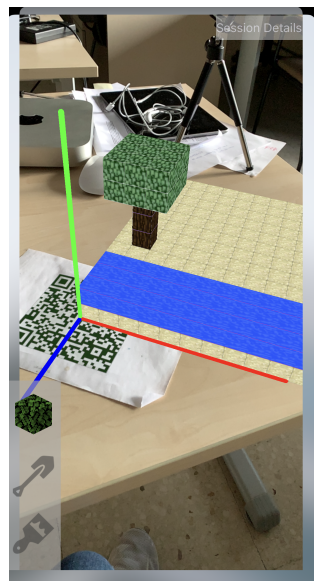

**Figure 14.** Capture of the voxel-based prototype taken from the iOS client.

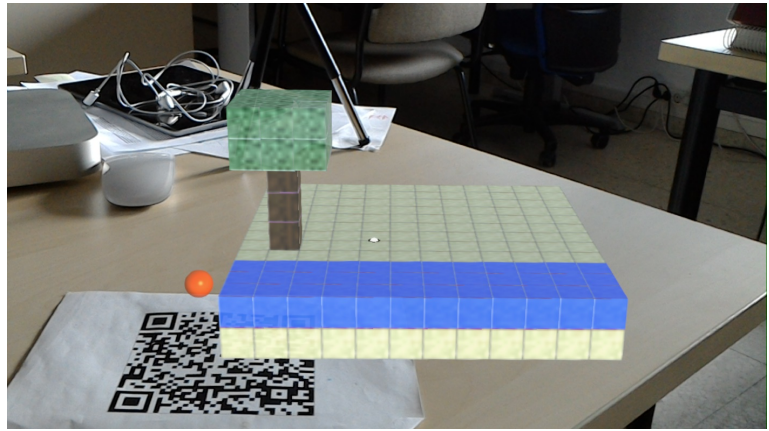

**Figure 15.** Capture of the voxel-based prototype taken from the HoloLens client.

5.  **Selecting and applying the *Collaboration Model*:** the fifth and final step is to establish the *Collaboration Model* that the session will present. In the voxels example, several possibilities are viable. For example, it is possible to use a *Turn-based* model, which will only allow the user who has the turn to make changes on the world. Another possibility is to limit the interaction capabilities of the users, allowing only one user to add cubes and another one to only delete them. However, it is also possible to set aside any rule by choosing the *Unconstrained model*. In that case, all participants may add and remove cubes from the world at their will, in a manner similar to the original version of the game where no control was performed.

In summary, in step 1 we launch the *Communication Service* which will allow the rest of the system to work. In step 2 we implement the application logic and prepare the connection with the SARA platform. With step 3, users' clients are generated, through which they will be able to see the AR content that represents the session status. In step 4, if required, the specific logic of the users' clients is implemented, which will depend on the scope of the application. Finally, in step 5 we select a Collaboration Model for the session that will make the appropriate regulation of the events of the session. This prototype can be seen in action in the following link [37]. As shown in Figure 6, there are three stages that are common among multiple developments over SARA: the configuration and launching of the *Communication Service* (step 1), the creation of the consumer clients (step 3) and the selection of the Collaboration Model (step 5). Thus, SARA provides artifacts with the basic functionality associated with each of these stages, which supports the application development by reusing them.

## 7. Conclusions and Future Work

In this paper we presented SARA, a cross-platform microservice-based architecture for collaborative Augmented and Mixed Reality that aims at supporting the deployment of these type of services. SARA also allows the adaptation of applications not intended for collaborative AR, so that they present this functionality. In addition, it supports the management of collaboration models, allowing their reuse between applications with different functional objectives and setting up the schemes and data models required for integrating new ones. These advantages have been demonstrated by the development of a game prototype, the voxel-based world editor, in which a single-user game has been adapted to the collaborative AR context.

Further work is needed to make the architecture and its supporting tools usable in generalized scenarios.

Firstly, it is required to address the specific issues related to performance, which may be critical in large-scale applications. In this moment, communication between clients is done through the exchange of events encoded in JSON format. Thus, a first step toward enhancing the performance will be to dispense with JSON-based verbose encoding and use a binary format that allows compression.

This will reduce the bandwidth required to send the messages, also reducing their size and alleviating JSON parsing times (which generally tend to be high).

Second, when defining both the Meshes and the Materials of the AR assets, for the sake of clarity it was decided to keep them simple. However, most 3D editing programs allow, for example, grouping several meshes under one or applying several materials and textures to the same object. The incorporation of these improvements in SARA would not be a problem, since as discussed in Section 5, the data model can be easily extended.

Third, a very common practice in 3D development environments is the use of *Shaders*. A Shader is a program fragment in charge of representing the visual elements and what is their appearance. This includes, for example, lights and shadows, as well as more advanced effects such as brightness, emission and reflections. In the practice of these environments, it is possible to change the visual appearance of an element without having to change the associated mesh, i.e., using one of those shaders. They are also used to "draw" visual elements that would otherwise consume many resources or calculations, such as particle systems. The incorporation of shaders in SARA will improve its graphic section and homogenize the representation of the elements among all platforms.

Fourth and finally, the integration of sound and animations is a highly desired feature. The vast majority of application used every day make use of sound or music effects to give feedback to the user or to improve their experience. With the implementation of SARA presented in this paper, the only way to integrate sound is to have predefined which are the audio files to play. Then, when a certain event is received in the *consumer* clients, these audio files are played. However, this approach is not flexible. For its part, the animation of 3D elements is also essential in many applications, especially in video games. SARA is able to handle changes in the basic characteristics of the nodes of a session (e.g., position, rotation and mesh information) and when they are animated or changed, participants will be able to see that animation as well. It would be desirable to analyze SARA's behavior with advanced animations as well as the integration of these animations themselves.

Thus, we have organized future technical work around (1) performance, (2) representation issues (meshes, materials and shaders) and (3) integration of sound and video. As the reader may notice, these are challenging aspects but are not affecting the general philosophy around collaboration models that drives the architecture, which may be extended and adapted to new service scenarios that may require different collaboration schemes.

Apart from considering these technical challenges, empirical evaluation on SARA perceived value and impact on real development practices could be addressed over a stable and fully functional implementation of the architecture. This would also involve recruiting a sample of developers with comparable skills to complete specific tasks over the framework. This approach could help to better assess the benefits and unveil limitations of the architecture. Our current further work is focused on exploiting SARA as a technology layer for prototyping applications to drive user studies within different service domains, technology settings and collaboration scenarios (based on a single model or combining several ones). Hopefully, these experiences will allow analyzing how the proposed collaboration models fit within mental models and to which extent the practical implementation of the proposed collaboration is acceptable in a realistic setting (with device, interaction and real time constraints) for a wider population. An additional issue to explore are the similarities and differences that remote and local contexts pose on collaboration and the possibility of delivering cross-domain collaborative AR interaction models.

**Author Contributions:** D.V.-M. and A.M.B. contributed the work concept, the design of the architecture and the writing. D.V.-M. implemented the technical layer with support of L.B. All authors have read and agreed to the published version of the manuscript

**Funding:** This work was supported by UPM Project RP150955017, and by the Spanish Ministry of Economy and Competitiveness under grant TEC2017-88048-C2-1-R.

**Conflicts of Interest:** The authors declare no conflict of interest.